%% file: hallac-kdd-2015.tex
\documentclass{sig-alternate-2013} 
\usepackage{color}
\usepackage{url}
\usepackage{verbatim} 
\usepackage{subfigure} 
\usepackage{multirow}
\usepackage{algorithm}
\usepackage[noend]{algorithmic}
\usepackage{xspace}
\usepackage{graphicx}
\usepackage{epsfig}
\usepackage{amsmath}
\usepackage{amssymb}
\usepackage{times}
\usepackage{enumitem}

\usepackage{dcolumn}

\usepackage{graphicx,psfrag,amsmath,amsfonts,appendix}
\usepackage{algorithm}

\newcommand{\reals}{{\mbox{\bf R}}}

\newcommand{\argmin}{\mathop{\rm argmin}}

\newcommand{\hide}[1]{}
\newcommand{\xhdr}[1]{\vspace{1.7mm}\noindent{{\bf #1.}}}

\newcommand{\ie}{\emph{i.e.}}

\graphicspath{{./FIG/}}

\newfont{\mycrnotice}{ptmr8t at 7pt}
\newfont{\myconfname}{ptmri8t at 7pt}
\let\crnotice\mycrnotice%
\let\confname\myconfname%

\permission{Permission to make digital or hard copies of all or part of this work for personal or classroom use is granted without fee provided that copies are not made or distributed for profit or commercial advantage and that copies bear this notice and the full citation on the first page. Copyrights for components of this work owned by others than the author(s) must be honored. Abstracting with credit is permitted. To copy otherwise, or republish, to post on servers or to redistribute to lists, requires prior specific permission and/or a fee. Request permissions from Permissions@acm.org.}
\conferenceinfo{KDD'15,}{August 10-13, 2015, Sydney, NSW, Australia. \\ 
{\mycrnotice{Copyright is held by the owner/author(s). Publication rights licensed to ACM.}}}
\copyrightetc{ACM \the\acmcopyr}
\crdata{978-1-4503-3664-2/15/08\ ...\$15.00.\\
DOI: http://dx.doi.org/10.1145/2783258.2783313 }

\begin{document}

\title{
Network Lasso: Clustering and Optimization in Large Graphs}

\numberofauthors{1}
\author{
\alignauthor David Hallac, Jure Leskovec, Stephen Boyd \\
        \affaddr{Stanford University} \\
        \email{\{hallac, jure, boyd\}@stanford.edu} \\
}

\maketitle

\begin{abstract}
\input{000abstract}
\end{abstract}

\vspace{1mm}
 \noindent {\bf Categories and Subject Descriptors:} H.2.8 {\bf
[Database Management]}: Database applications---{\it Data mining}

\noindent {\bf General Terms:} Algorithms; Experimentation.


\section{Introduction}
\label{sec:intro}
\input{010intro}

\section{Convex Problem Definition}
\label{sec:original}
\input{020original}

\section{Proposed Solution}
\label{sec:proposed}
\input{030proposed}

\section{Non-Convex Extension}
\label{sec:nonconvex}
\input{040nonconvex}


\section{Experiments}
\label{sec:experiments}
\input{060experiments}

\section{Conclusion and Future Work}
\label{sec:conclusion}
\input{080conclusion}

\bibliography{refs}
\bibliographystyle{abbrv}

\appendix

\section{Analytical Solution to $z$-update}
\label{sec:appendixB}
\input{100appendixB}

\end{document}

%% file: 000abstract.tex
Convex optimization is an essential tool for modern data analysis, as it provides a framework to formulate and solve many problems in machine learning and data mining. However, general convex optimization solvers do not scale well, and scalable solvers are often specialized to only work on a narrow class of problems. Therefore, there is a need for simple, scalable algorithms that can solve many common optimization problems. In this paper, we introduce the \emph{network lasso}, a generalization of the group lasso to a network setting that allows for simultaneous clustering and optimization on graphs. We develop an algorithm based on the Alternating Direction Method of Multipliers (ADMM) to solve this problem in a distributed and scalable manner, which allows for guaranteed global convergence even on large graphs. We also examine a non-convex extension of this approach. We then demonstrate that many types of problems can be expressed in our framework. We focus on three in particular --- binary classification, predicting housing prices, and event detection in time series data --- comparing the network lasso to baseline approaches and showing that it is both a fast and accurate method of solving large optimization problems. 

%% file: 010intro.tex
Convex optimization has become an increasingly popular way of modeling problems in many different fields, ranging from 
finance \cite[\S 4.4]{BV:04} to image processing \cite{CRT:06}.
However, as datasets get larger and more intricate, classical methods of convex analysis, which often rely on interior point methods, begin to fail due to a lack of scalability. In fact, without any known structure to the optimization problem, the convergence time will scale with the cube of the problem size \cite{BV:04}.
The challenge of large-scale optimization lies in developing methods general enough to work well independent of the input and capable of scaling to the immense datasets that today's applications require. Presently, solving these problems in a scalable way requires
developing problem-specific solvers to
exploit structure in the model \cite{YMW:06}, often an infeasible assumption. 
Therefore, it is necessary to formulate general classes of optimization solvers that can apply to a variety of 
relevant problems, and to develop algorithms for obtaining reliable and efficient solutions.

\xhdr{Present Work: Formulation}
Here, we focus on optimization problems posed on graphs. Consider the following problem on a graph 
$\mathcal G=(\mathcal V, \mathcal E)$, where $\mathcal{V}$ is the vertex set and $\mathcal{E}$ the set of edges:
\begin{equation} 
\begin{array}{ll} \mbox{minimize} & \sum\limits_{i \in \mathcal V} f_i(x_i)
+ \sum\limits_{(j,k)\in \mathcal E} g_{jk}(x_j,x_k).
\end{array}
\end{equation}
The variables are $x_1, \ldots, x_m \in \reals^p$, where $m=| \mathcal V|$.
(The total number of scalar variables is $mp$.)
Here $x_i\in \reals^p$ is the variable at node $i$, $f_i:\reals^p \to
\reals \cup \{\infty\}$ is the cost function at node $i$,
and $g_{jk}: \reals^p \times \reals^p \to \reals \cup \{\infty\}$
is the cost function associated with edge $(j,k)$.
We use extended (infinite) values of $f_i$ and $g_{jk}$ to 
describe constraints on the variables, or pairs of variables 
across an edge, respectively.
Our focus will be on the special case in which the $f_i$ are convex,
and $g_{jk}(x_j,x_k) = \lambda w_{jk} \|x_j-x_k\|_2$,
with $\lambda \geq 0$ and user-defined weights $w_{jk} \geq 0$:
\begin{equation}\label{original}
\begin{array}{ll} \mbox{minimize} & \sum\limits_{i \in \mathcal V} f_i(x_i)
+ \lambda \sum\limits_{(j,k)\in \mathcal E} w_{jk}\|x_j-x_k\|_2.
\end{array}
\end{equation}
The edge objectives penalize differences between the variables
at adjacent nodes, where the edge between nodes $i$ and $j$ has weight $\lambda w_{ij}$.
Here we can think of $w_{ij}$ as setting the relative weights
among the edges of the network, and $\lambda$ as an overall 
parameter that scales the edge objectives relative to the node
objectives.
We call problem \eqref{original} the \emph{network lasso} problem, since the edge cost
is a sum of norms of differences of the adjacent edge variables. 

The network lasso problem
is a convex optimization problem, and so in principle it can be solved
efficiently.   For small networks, generic (centralized) 
convex optimization methods can be used to solve it.
But we are interested in problems with many variables, with $p$, $m=| \mathcal V|$, and $n=| \mathcal E|$ all potentially large. For such problems no adequate solver currently exists.
Thus, we develop a distributed and scalable
method for solving the network lasso problem, in which each vertex
variable $x_i$ is controlled by one ``agent'', and the agents 
exchange (small) messages
over the graph to solve the problem iteratively. This approach 
provides global convergence for all problems that can be put into this form.
We also analyze a non-convex extension of the network lasso, a slightly different way to model the problem, and
give a similar algorithm that, although it does not guarantee optimality, 
tends to perform well in practice.

\xhdr{Present Work: Applications}
There are many general settings in which the network lasso
problem arises. In control systems, the nodes might represent the possible states of a system,
and $x_i$ the action or actions to take when we are in state $i$,
so the collection of variables $(x_1,\ldots, x_m)$ describes a policy.
The graph tells us about state transitions, and the weights 
express how much we care about the actions in neighboring states 
differing.
Here the network lasso problem seeks a solution that
minimizes the total cost, but also does not change much 
across adjacent states, allowing for a ``simpler'' policy.
The parameter $\lambda$ allows us to trade off the total cost
(the node objective) versus a cost for the actions 
varying across the states (the edge objective).

Another general setting, one we focus on in this paper, relates to statistical learning, where the variables $x_i$
are parameters in a statistical model of some data resident at, or associated with,
node $i$.  
The objective term $f_i$ represents the loss for the model
over the data, possibly with some regularization added in.
The edge terms are regularization that encourages adjacent nodes
to have close (or the same) model parameters.
In this setting, the network expresses our
idea that adjacent nodes should have similar (or the same) models.
We can imagine that this regularization allows us to build models 
at each node that borrow strength from the fact
that neighboring nodes should have similar, or even identical, models.

It is critical to note that the edge terms in the network lasso problem involve the norm, not the norm squared, of 
the difference. If the norms were
squared, the edge objective would reduce to (weighted) Laplacian
regularization \cite{WSZS:06}.
The sum-of-norms regularization that we use is like group lasso
\cite{YL:06}; it encourages not just $x_i\approx x_j$ for edge $(i,j)\in
\mathcal E$, but $x_i=x_j$, \ie, 
\emph{consensus} across the edge.
Indeed, we will see that there is often a (finite) value of 
$\lambda$ above which the solution 
has $x_1=\cdots=x_m$, \ie, all the vectors are in
consensus.
For smaller values of $\lambda$, the solution of the 
network lasso problem breaks into clusters of 
nodes, with $x_i$ the same across all nodes in the cluster.
In the policy setting, we can think of this as a combination of 
state aggregation or clustering, together with policy design.
In the modeling setting, this is a combination of 
clustering the data collections and fitting a model to each cluster.

\xhdr{Present Work: Use Case}
As a running example, which we later analyze in detail, consider the problem of predicting housing prices. 
One common approach is linear regression. That is, we learn the weights of each feature (number of bedrooms, square footage, etc...) and use these same weights for each house to estimate the price.
However, due to location-based factors such as school district or distance to a highway, similar houses in different locations can have drastically different prices. These factors are often unknown a priori and difficult to quantify, so it is inconvenient to attempt to incorporate them as features in the regression.
Therefore, standard linear regression will have large errors in price prediction, since it forces the entire dataset to agree on a single global model. 
What we actually want is to cluster the houses into ``neighborhoods'' which share a common regression model.
First, we build a network where neighboring houses (nodes) are connected by edges. Then, each house solves for its own regression model (based on its own features and price). We use the network lasso penalty to encourage nearby houses to share the same regression parameters, in essence helping each house determine which neighborhood it is part of, and learning relevant information from this group of neighbors to improve its own prediction.
The size and shape of these neighborhoods, though, are difficult to know beforehand
and often depend on a variety of factors, including the amount of available data.
The network lasso solution empirically determines the neighborhoods,
so that each house can share a common model with houses in its cluster,
without having to agree with the potentially misleading information from other locations.

\xhdr{Summary of Contributions}
The main contributions of this paper are as follows:

\begin{itemize}[nolistsep]

\item We formally define the \emph{network lasso}, a specific type of optimization problem on networks.

\item We develop a fast, scalable, and distributed solver for any problem of this form. This algorithm is also capable of choosing the right regularization parameter $\lambda$.

\item We show that many common and useful problems can be
formulated as an instance of the network lasso.

\end{itemize}

\xhdr{Related Work}
The network lasso can be thought of as a special case of certain methods (Bayesian inference, general convex optimization) and a generalization of others (fused lasso \cite{TS:05}, total variation \cite{WBAW:12, YWFZWY:13}). 
It occupies a unique point on the trade-off curve between generality and scalability that, to the best of our knowledge, has not yet been formally analyzed. Our approach provides a unified view of a diverse class of optimization problems, but is still capable of solving large-scale examples. For example, convex clustering \cite{CL:13, HJBV:11, PDSD:05}, an alternative to the K-means algorithm, is a well-studied instance of the network lasso. However, convex clustering requires $f_i$ to be the square loss from some observation $a_i$, and often assumes a fully connected graph since there is no prior information about which nodes may be clustered together. In contrast, generalizing to any shape of network with any convex objectives (including allowing constraints) allows our approach to be applied to new topics, such as control systems and event detection. Furthermore, we elect to focus on the $\ell_2$-norm because of its intuitive network-based rationale in that it leads to node stratification.

The network lasso is also related to probabilistic graphical models (PGMs). Problem \eqref{original} is a type of Bayesian inference where we learn a set of models or dependencies based on latent clustering.
The network lasso penalty, a form of regularization, allows for one type of ``relationship'' between nodes, a weighted prior belief that the connected variables should be equal. 
The clustering that our model accomplishes is similar to a latent variable mixture model \cite{M:01}, where cluster membership is indicated by some latent variable. With this, certain network lasso problems can be rewritten as a maximum likelihood estimation problem where a conditional distribution is learned for each cluster. However, many examples are difficult to encode and scale in this way. 
Additionally, there has been much research on optimal decomposition and splitting methods for these types of problems \cite{CLCD:07, MJ:01}.
Hinge-loss Markov random fields, which are PGMs defined over continuous variables for MAP inference, use a similar ADMM-based approach to ours \cite{BHLG:13}, though the hinge-loss potentials they support do not include the norm-based lasso that we utilize to induce the clustering.
However, unlike many of these other frameworks \cite{BHLG:13, KS:05, ZR:14}, which often use a probabilistic approach, we formulate it as a single, very large, convex optimization problem that we solve by splitting it across a graph. 
This focus on the specific topic of simultaneous clustering and optimization enables us to provide a clean formalism and scalable approach, with guaranteed convergence, for solving a wide class of problems, all using the exact same algorithm.

%% file: 020original.tex
We now look more closely at the network lasso problem,
\begin{equation*}
\begin{array}{ll} \mbox{minimize} & \sum\limits_{i \in \mathcal V} f_i(x_i)
+ \lambda \sum\limits_{(j,k)\in \mathcal E} w_{jk}\|x_j-x_k\|_2.
\end{array}
\end{equation*}
This problem is convex in the variable $x = (x_1, \ldots, x_m) \in \reals^{mp}$, and we let $x^{\star}$ denote an optimal solution. 

\xhdr{Local Variables}
It is worth noting that there can be local private optimization variables at each node that are not part of the lasso penalty. More formally, the network lasso problem can be defined as 
\begin{equation}\label{formal-net-lasso}
\begin{array}{ll} \mbox{minimize} & \sum\limits_{i \in \mathcal V} \tilde{f}_i(x_i, \varepsilon_i)
+ \lambda \sum\limits_{(j,k)\in \mathcal E} w_{jk}\|x_j-x_k\|_2,
\end{array}
\end{equation}
where $\varepsilon_i$ are potential dummy variables at node $i$ (the size can vary at each node).
However, using partial minimization, if we let
\begin{equation*}
\begin{array}{ll} f_i(x_i) = \min\limits_{\varepsilon_i} \tilde{f}_i(x_i, \varepsilon_i),
\end{array}
\end{equation*}
we get the original problem, defined in (\ref{original}). For simplicity, we therefore use problem \eqref{original} throughout the paper, with the implicit understanding that there may be private variables at each node.


\xhdr{Regularization Path}
Although the regularization parameter $\lambda$ in problem \eqref{original} can be incorporated into the $w_{ij}$'s by scaling the edge weights, it is best viewed separately as a single parameter which is tuned to yield different global results. $\lambda$ defines a trade-off for the nodes between minimizing its own objective and agreeing with its neighbors. At $\lambda = 0$, $x_i^\star$, the solution at node $i$, is simply a minimizer of $f_i$. This can be computed locally at each node, since when $\lambda=0$ the edges of the network have no effect. At the other extreme, as $\lambda \rightarrow \infty$, problem \eqref{original} turns into
    \begin{equation}
        \begin{array}{ll}
           \mbox{minimize}  &\sum\limits_{i \in \mathcal{V}}f_{i}(\tilde{x}),
        \end{array}
        \label{GenCon}
        \end{equation}
since a common $\tilde{x}$ must be the solution at every node. This is solved by $x^{\mathrm{cons}} \in \reals^p$. We refer to \eqref{GenCon} as the \emph{consensus problem} and to $x^{\mathrm{cons}}$ as the \emph{consensus solution}. If a solution to \eqref{GenCon} exists, it can be shown that there is a finite $\lambda_{\mathrm{critical}}$ such that for any $\lambda \geq \lambda_{\mathrm{critical}}$, the consensus solution holds. That is, beyond this $\lambda_{\mathrm{critical}}$, increasing $\lambda$ has no effect on the solution. For $\lambda$'s in between $\lambda = 0$ and $\lambda_{\mathrm{critical}}$, the family of solutions is known as the \emph{regularization path}, though it is sometimes referred to as the clusterpath \cite{HJBV:11}.


\xhdr{Network Lasso and Clustering} 
The $\ell_2$-norm penalty over the edge difference, $\|x_{j} - x_{k}\|_{2}$, defines the \emph{network lasso}. It incentivizes the differences between connected nodes to be exactly zero, rather than just close to zero, yet it does not penalize large outliers (in this case, node values being very different) too severely. An edge difference of zero means that $x_j = x_k$.  When many edges are in consensus like this, we have grouped the nodes into sets with equal values of $x_i$. Each set of nodes, or cluster, has a common solution for the variable $x_i$. The outliers then refer to edges between nodes in different clusters. Cluster size tends to get larger as $\lambda$ increases, until at $\lambda_{\mathrm{critical}}$ the consensus solution can be thought of as a single cluster for the entire network. Even though increasing $\lambda$ is most often agglomerative, cluster fission may occur, meaning two nodes in the same cluster may break apart at a higher $\lambda$. Therefore, the clustering pattern is not strictly hierarchical \cite{PDSD:05}. 


\xhdr{Inference on New Nodes}
After we have solved for $x^{\star}$, we can interpolate the solution to estimate the value of $x_j$ on a new node $j$, for example during cross-validation on a test set.
Given $j$, all we need is its location within the network; that is, the neighbors of $j$ and the edge weights. With this information, we treat $j$ like a dummy node, with $f_j(x_j) = 0$. We solve for $x_j$ just like in problem \eqref{original} except without the objective function $f_j$, so the optimization problem becomes
      \begin{equation} 
        \begin{array}{ll}
           \mbox{minimize}  &\sum\limits_{k \in N(j)}w_{jk}\|x_j - x_k^\star\|_2,
        \end{array}
        \label{inference}
       \end{equation}
where $N(j)$ is the set of neighbors of node $j$. This estimate of $x_j$ can be thought of as a weighted median of $j$'s neighbors' solutions. This is called the Weber problem, and it involves finding the point which minimizes the weighted sum of distances to a set of other points \cite{BMM:03}. It has no analytical solution when $j$ has more than two neighbors, but it can be readily computed even for large problems. 
For example, when one of the dimensions is much larger than the other (number of neighbors vs.\ size of each $x_k$), the problem can be solved in linear time with respect to the larger dimension \cite{BV:04}.

%% file: 030proposed.tex
On smaller graphs, the network lasso problem can be solved using standard interior point methods. This paper focuses on large problems, where solving everything at once is infeasible. This is especially true when we solve for a span of $\lambda$'s across the entire regularization path, since we will need to solve a separate problem for each $\lambda$.
A distributed solution is necessary so that computational and storage limits do not constrain the scope of potential applications. We propose an easy-to-implement algorithm based on the Alternating Direction Method of Multipliers (ADMM) \cite{BPCPE:11, PB:14}, a well-established method for distributed convex optimization. With ADMM, each individual component solves its own private objective function, passes this solution to its neighbors, and repeats the process until the entire network converges. There is no need for global coordination except for iteration synchronization.

\subsection{ADMM}
To solve via ADMM, we introduce a copy of $x_i$, called $z_{ij}$, at each edge $ij$. Note that the same edge also has a $z_{ji}$, a copy of $x_j$. We rewrite problem $\eqref{original}$ as an equivalent problem,
      \begin{equation*} 
        \begin{array}{ll}
           \mbox{minimize}  &\sum\limits_{i \in \mathcal{V}}f_{i}(x_{i}) + \lambda \sum\limits_{(j,k)\in \mathcal{E}} w_{jk}\|z_{jk} - z_{kj}\|_{2} \\
           \mbox{subject to} &x_i = z_{ij}, \quad i = 1,\ldots,m, \quad j \in N(i).
        \end{array}
       \end{equation*}
We then derive its augmented Lagrangian \cite{H:69}, which gives us
    \begin{align*}
              L_\rho(x,z,u) = &\sum\limits_{i \in \mathcal{V}}f_{i}(x_{i}) + \sum\limits_{(j,k)\in \mathcal{E}} \biggl(  \lambda w_{jk}\|z_{jk} - z_{kj}\|_{2} - \\
              & (\rho/2) \left(\|u_{jk}\|_2^2 + \|u_{kj}\|_2^2\right)+ \\
              &  (\rho/2)\left(\|x_j - z_{jk} + u_{jk}\|_2^2 + \|x_k - z_{kj} + u_{kj}\|_2^2\right) \biggr),
      \end{align*}
where $u$ is the scaled dual variable and $\rho > 0$ is the penalty parameter \cite[\S 3.1.1]{BPCPE:11}. ADMM consists of the following steps, with $k$ denoting the iteration number:
       \begin{align*}
           &x^{k+1} = \argmin_x L_\rho(x,z^k,u^k) \\
           &z^{k+1} = \argmin_z L_\rho(x^{k+1},z,u^k) \\
           &u^{k+1} = u^{k} + (x^{k+1} - z^{k+1}).
        \end{align*}       
Let us examine each of these steps in more detail.

\xhdr{$x$-Update} 
 In the $x$-update we minimize a separable sum of functions, one per node,
so it can be calculated independently at each node and solved in parallel.
At node $i$, this is
         \begin{equation*}
        x_i^{k+1} =  \underset{x_i}{\mathrm{argmin}} \left(f_i(x_i) + \sum\limits_{j \in N(i)} (\rho/2)\|x_i - z_{ij}^k + u_{ij}^k\|_2^2\right).
        \end{equation*}
        
\xhdr{$z$-Update}
The $z$-update is separable across the edges. Note that for edge $ij$, 
we need to jointly update $z_{ij}$ and $z_{ji}$. This becomes 
         \begin{align*}
     z_{ij}^{k+1}, z_{ji}^{k+1} = &\underset{z_{ij}, z_{ji}}{\mathrm{argmin}} \biggl(\lambda w_{ij}\|z_{ij} - z_{ji}\|_2 + \\
     & (\rho/2)\left(  \|x_i^{k+1} - z_{ij} + u_{ij}^k\|_2^2 + \|x_j^{k+1} - z_{ji} + u_{ji}^k\|_2^2 \right)  \biggr).
        \end{align*}
This problem has a closed-form analytical solution, which we derive in Appendix A.
It is
         \begin{align*}
     &z_{ij}^\star = \theta(x_i + u_{ij}) + (1-\theta)(x_j + u_{ji}) \\
       & z_{ji}^\star = (1-\theta)(x_i + u_{ij}) + \theta(x_j + u_{ji}),       
        \end{align*}
where
         \begin{equation}\label{thetaDef}
     \theta = \mathrm{max}\left(1 - \frac{\lambda w_{ij}}{\rho \|x_i + u_{ij} - (x_j + u_{ji})\|_2}, 0.5 \right). 
        \end{equation}

\xhdr{$u$-Update}
The $u$-update is also edge-separable. For each variable, this looks like 
         \begin{equation*}
        u_{ij}^{k+1} = u_{ij}^{k} + (x_i^{k+1} - z_{ij}^{k+1}).          
        \end{equation*}

\xhdr{Global Convergence}       
Because the problem is convex, ADMM is guaranteed to converge to a global optimum. The stopping criterion can be based on the primal and dual residuals, commonly defined as $r$ and $s$, being below given threshold values; see \cite{BPCPE:11}.  
This allows us to stop when $x^k$ and $z^k$ are close, and when $x^k$ (or $z^k$) does not change much in one iteration. As is typical for ADMM, the algorithm tends to attain modest accuracy relatively quickly, and high accuracy (which in many applications is not needed) only slowly.
\begin{algorithm}
\caption{ADMM Steps}\label{ADMMiterations}
\begin{tabbing}
    {\bf repeat} \\
    \qquad \= $x_i^{k+1} =  \underset{x_i}{\mathrm{argmin}} \left(f_i(x_i) + \sum\limits_{j \in N(i)} (\rho/2)\|x_i - z_{ij}^k + u_{ij}^k\|_2^2\right)$ \\
    \> $z_{ij}^{k+1} = \theta(x_i + u_{ij}) + (1-\theta)(x_j + u_{ji})$ \\
    \> $z_{ji}^{k+1} = (1-\theta)(x_i + u_{ij}) + \theta(x_j + u_{ji})$ \\   
    \> $u_{ij}^{k+1} = u_{ij}^{k} + (x_i^{k+1} - z_{ij}^{k+1})$ \\*[\smallskipamount]
    {\bf until} \= $\|r^k\|_2 \leq \epsilon^{\mathrm{pri}}$; $\|s^k\|_2 \leq \epsilon^{\mathrm{dual}}$.
\end{tabbing}
\end{algorithm}

\subsection{Regularization Path}
It is often useful to compute the regularization path as a function of $\lambda$ to gain insight into the network structure. For specific applications, this may also help decide the correct value of $\lambda$ to use, for example by choosing $\lambda$ to minimize the cross-validation error.

We begin the regularization path at $\lambda = 0$ and solve for an increasing sequence of $\lambda$'s ($\lambda := \alpha \lambda$, $\alpha > 1$). We know when we have reached $\lambda_{\mathrm{critical}}$ because a single $x^{\mathrm{cons}}$ will be the optimal solution at every node, and increasing $\lambda$ no longer affects the solution. This may lead to a stopping point slightly above the actual $\lambda_{\mathrm{critical}}$, which we denote as $\tilde{\lambda}_{\mathrm{critical}}$. There is no harm if $\tilde{\lambda}_{\mathrm{critical}} > \lambda_{\mathrm{critical}}$, since they will both yield the same result, the consensus solution. To account for the case where no consensus solution exists, we can also stop when the new solution has changed by less than some $\epsilon$, since even without consensus, the problem converges to some solution.

A big advantage of the regularization path, as opposed to computing each value of $x^\star(\lambda)$ in parallel, is that we begin with a warm start towards the new solution at each step. For each $\lambda$, the unknown variables are already close to the new $x^\star$, $u^\star$, and $z^\star$ by virtue of starting at the solution for the last $\lambda$. In fact, when $f_i$ is strictly convex, the solution $x^\star$ is continuous in $\lambda$. Without any prior knowledge, for example initializing everything to 0 for each $\lambda$, we start far from the actual solution, so it will often (although not always) take longer to converge via ADMM. The only other required variable is $\lambda_{\mathrm{initial}}$, the initial non-zero value of $\lambda$, which depends on the variable scaling. The hope is that $x^\star$ does not change too much between $\lambda = 0$ and this initial value, and a rough estimate of $\lambda_{\mathrm{initial}}$ can be found using the following heuristic:
\begin{enumerate}[nolistsep]
  \item Pick edge $ij$ at random and find $x_i^\star$, $x_j^\star$ at $\lambda = 0$.
  \item Evaluate the gradients of $f_i(x)$ and $f_j(x)$ at $x = (x_i^\star + x_j^\star)/2$.
  \item Set $\lambda_{\mathrm{initial}} := 0.01 \left(\frac{\|\nabla f_i(x)\|_2 + \|\nabla f_j(x)\|_2}{2 w_{ij}}\right)$.
\end{enumerate}
To get a more robust estimate, repeat the above steps picking different edges each time, and choose the smallest solution for $\lambda_{\mathrm{initial}}$. Given these variables, we are now able to solve for the entire regularization path. Our method is outlined in Algorithm \ref{RegPathAlgo}.

\begin{algorithm}
\caption{Regularization Path}\label{RegPathAlgo}
\begin{tabbing}
    {\bf initialize} Solve for $x^\star$, $u^\star$, $z^\star$ at $\lambda = 0$. \\*[\smallskipamount]
    {\bf set} $\lambda := \lambda_{\mathrm{initial}}$; $\alpha > 1$; $u := u^\star$; $z := z^\star$.\\*[\smallskipamount]
    {\bf repeat} \\
    \qquad \= Use ADMM to solve for $x^\star(\lambda)$ (see Algorithm \ref{ADMMiterations}) \\
    \> $\textit{Stopping Criterion.}$ $\textbf{quit}$ if $x^\star(\lambda) = x^\star(\lambda_{\mathrm{previous}})$ \\
    \> Set $\lambda := \alpha \lambda$.\\*[\smallskipamount]
    {\bf return} $x^\star(\lambda)$ for $\lambda$ from $0$ to $\tilde{\lambda}_{\mathrm{critical}}$.
\end{tabbing}
\end{algorithm}

%% file: 040nonconvex.tex
\begin{figure}
    \centering
    \includegraphics[width=0.85\linewidth]{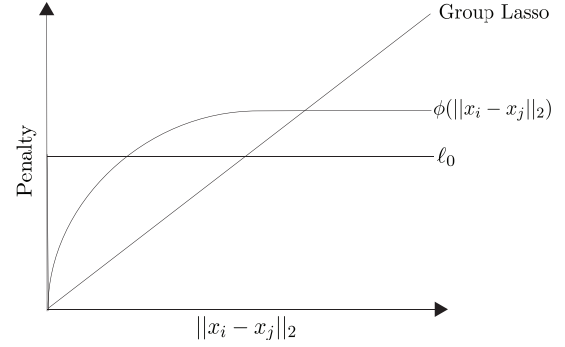}
    \vspace{-2mm}
    \caption{Comparison of Group Lasso, $\ell_0$, and Non-Convex $\phi$.}
    \vspace{-2mm}
    \label{fig:concave}
\end{figure}

In many applications, we are using the group lasso as an approximation of the $\ell_0$-norm \cite{CWB:08}. That is, we are looking for a sparse solution where relatively few edge differences are non-zero. However, once $\|x_i - x_j\|_2$ becomes non-zero, we do not care about its magnitude, since we already know that $i$ and $j$ are in different clusters. The lasso has a proportional penalty, which is the closest that a convex 
function can come to approximating the $\ell_0$-norm. Once we have found the true clusters, though, this will ``pull'' the different clusters towards each other through their mutual edges. If we replace the group lasso penalty with a monotonically nondecreasing concave function $\phi(u)$, where $\phi(0) = 0$ and whose domain is $u \geq 0$, we come even closer to the $\ell_0$, as shown in Figure \ref{fig:concave}.
However, this new optimization problem,
    \begin{equation}
        \begin{array}{ll}
           \mbox{minimize}  &\sum\limits_{i \in \mathcal{V}}f_{i}(x_{i}) + \lambda\sum\limits_{(j,k)\in\mathcal{E}} w_{jk} \phi \left( \|x_{j} - x_{k}\|_{2}\right),
        \end{array}
        \label{original2}
        \end{equation}
is not convex. ADMM is not guaranteed to converge, and even if it does, it need not be to a global optimum. It is in some sense a ``riskier'' approach. In fact, different initial conditions on $x$, $u$, $z$, and $\rho$ can yield quite different solutions. However, as a heuristic, a slight modification to ADMM empirically performs very well. Since the algorithm might not converge, it is necessary to keep track of the iteration which yields the minimum objective, and to return that as the solution instead of the most recent step. The primal and dual residuals are not guaranteed to go to $0$, so we instead run our algorithm for a set number of iterations for each $\lambda$.

\xhdr{Non-Convex $z$-Update} Compared to the convex case, the only difference in the ADMM solution is the $z$-update, which is now
         \begin{equation}
        \begin{array}{ll} 
     \mbox{minimize}  &\lambda w_{ij} \phi \left( \|z_{ij} - z_{ji}\|_2 \right) + (\rho/2) \bigl(  \|x_i^{k+1} - z_{ij} + u_{ij}^k\|_2^2 + \\
     &\|x_j^{k+1} - z_{ji} + u_{ji}^k\|_2^2 \bigr).
        \end{array}
        \label{nonconvexZ}
        \end{equation}
For simplicity, we define
         \begin{align*} 
     	a = x_i^{k+1} + u_{ij}^k, & \quad
	b = x_j^{k+1} + u_{ji}^k,\\
	    c = \lambda w_{ij}, &\quad d = \| a - b \|_2,
        \end{align*}
so problem \eqref{nonconvexZ} turns into
         \begin{equation*}
        \begin{array}{ll} 
      \mbox{minimize}  &c \phi \left( \|z_{ij} - z_{ji}\|_2 \right) + (\rho/2) \left(  \|a - z_{ij}\|_2^2 + \|b- z_{ji}\|_2^2 \right).
        \end{array}
        \end{equation*}

There are two possible cases for the solution to problem \eqref{nonconvexZ}: $z_{ij}^\star = z_{ji}^\star$ or $z_{ij}^\star \neq z_{ji}^\star$. When the two solutions are identical, then $\phi \left( \|z_{ij} - z_{ji}\|_2 \right) = \phi (0) = 0$, so the only terms remaining are 
         \begin{equation*}
	(\rho/2) \left(  \|a - z_{ij}\|_2^2 + \|b - z_{ji}\|_2^2 \right).
        \end{equation*}
Minimizing over the constraint that $z_{ij} = z_{ji}$ yields $z_{ij}^\star = z_{ji}^\star = (1/2)(a + b)$ and an objective of $(\rho/4) \| a - b\|^2_2$.        
        
When the two solutions are not equal, $z_{ij}^\star$ and $z_{ji}^\star$ must lie on the line segment between $a$ and $b$. If $z_{ij}^\star$ and/or $z_{ji}^\star$ are not on the line segment, projecting them onto this segment is nonincreasing in $\phi\left(\|z_{ij} - z_{ji}\|_2\right)$ and decreasing in $(\rho/2) \left(  \|a - z_{ij}\|_2^2 + \|b- z_{ji}\|_2^2 \right)$,
so the total objective function is guaranteed to decrease. Therefore, we know that 
        \begin{align*} 
     &z_{ij}^\star = \theta_1 a + (1-\theta_1)b, \quad \theta_1 \in [0,1] \\
       &z_{ji}^\star = \theta_2 a + (1-\theta_2)b, \quad \theta_2 \in [0, 1]
        \end{align*} 
and that
        \begin{equation*}
     \|z_{ij}^\star - z_{ji}^\star\|_2 = \|a - b\|_2 \left( |\theta_1 - \theta_2| \right) = d  |\theta_1 - \theta_2|.
        \end{equation*}
Note that the solution for $z_{ij}^\star = z_{ji}^\star$ is just $\theta_1 = \theta_2 = \frac{1}{2}$. We also know that $\theta_1 \geq \theta_2$. If $\theta_1 < \theta_2$, we could swap $\theta_1$ and $\theta_2$ and $\phi\left(\|z_{ij} - z_{ji}\|_2\right)$ would remain constant, but the rest of the objective, $(\rho/2) \left(  \|a - z_{ij}\|_2^2 + \|b- z_{ji}\|_2^2 \right)$, 
would decrease. Therefore, we can rewrite the norm of the difference as
        \begin{equation*}
     \|z_{ij}^\star - z_{ji}^\star\|_2 = d (\theta_1 - \theta_2),
        \end{equation*}
and the objective becomes
         \begin{equation*}
        \begin{array}{ll} 
     c \phi \left( d (\theta_1 - \theta_2) \right) + (\rho d^2/2) \left( (1 - \theta_1)^2 + \theta_2^2 \right).
        \end{array}
        \end{equation*}
When $z_{ij}^\star \neq z_{ji}^\star$, we know that $\theta_1 > \theta_2$, and thus $d (\theta_1 - \theta_2) > 0$. When $\phi$ is differentiable at $d(\theta_1 - \theta_2)$, we set the gradient to zero:
         \begin{align*}
	&\frac{\partial }{\partial \theta_1} = c d \phi^\prime (d(\theta_1 - \theta_2)) - \rho d^2 (1 - \theta_1) = 0 \\
	&\frac{\partial }{\partial \theta_2} = -c d \phi^\prime (d(\theta_1 - \theta_2)) + \rho d^2 \theta_2 = 0.
        \end{align*}
We see that
         \begin{equation*}
	\rho d^2 (1 - \theta_1) = c d \phi^\prime (d(\theta_1 - \theta_2)) = \rho d^2 \theta_2,
        \end{equation*}
or
         \begin{equation*}
	\theta_2 = 1-\theta_1.
        \end{equation*}
This puts the entire optimization problem in terms of one variable, $\theta = \theta_2$. Since $\theta_1 + \theta_2 = 1$ and $\theta_1 \geq \theta_2$, we know that $\theta \leq \frac{1}{2}$, so the final problem becomes
         \begin{equation}
        \begin{array}{ll} 
	 \mbox{minimize}  &c \phi \left( d (1 - 2\theta) \right) + \rho d^2 \theta^2 \\
     \mbox{subject to} & 0 \leq \theta \leq \frac{1}{2}.
        \end{array}
        \label{1dProblem}
        \end{equation}
It is of course necessary to find all solutions to this problem, since there may be several or none, and to compare the resulting objective to $(\rho/4) \| a - b\|^2_2$, when $z_{ij}^\star = z_{ji}^\star$. Of these solutions, pick the $z$'s which minimize the overall objective function.

\xhdr{Log Function} We will now look at the specific case where $\phi(u) = \log(1 + \frac{u}{\epsilon})$, where $\epsilon$ is a constant scaling factor. The objective function in problem \eqref{1dProblem} turns into
         \begin{equation*}
        \begin{array}{ll} 
	 \mbox{minimize}  &c  \log(1 + \frac{d(1-2\theta)}{\epsilon}) + \rho d^2 \theta^2.
        \end{array}
        \end{equation*}
Setting the derivative equal to zero, we get 
         \begin{equation*}
	-\frac{2cd}{d - 2d\theta + \epsilon} + 2 \rho d^2 \theta = 0.
        \end{equation*}
We simplify to
         \begin{equation*}
	2\rho d^2\theta^2 - \rho d(d+\epsilon) \theta + c = 0
        \end{equation*}
and see that this is a simple quadratic equation in $\theta$, solved by
        \begin{equation*}
	\theta = \frac{\rho (d+\epsilon) \pm \sqrt{\rho^2 (d+\epsilon)^2 - 8 \rho c}}{4 \rho d}.
        \end{equation*}
The $z$-update then involves comparing the resulting objectives with $(\rho/4) \| a - b\|^2_2$ (the value when $z_{ij}^\star = z_{ji}^\star$) and then choosing the $\theta$ which yields the best of the three objectives to obtain $z_{ij}^\star$, $z_{ji}^\star$. If the quadratic term has no real roots, which happens more frequently as $\lambda$ increases, we set $\theta = \frac{1}{2}$, meaning the edge is in consensus. It is worth reiterating that this method is not guaranteed to reach the global optimum. Instead, it is an easy-to-implement algorithm that parallels ADMM from the convex case. 

%% file: 060experiments.tex
We now apply our approach on three examples to illustrate the diverse set of problems that fall under the network lasso framework, and to provide a simple and unified view of these seemingly different applications.
First, we look at a synthetic example in which we gather statistical power from the network to improve classification accuracy.
Next, we see how our approach can apply to a geographic network, allowing us to gain insights on residential neighborhoods by predicting housing prices. Finally, we look at a time series dataset for the purpose of detecting outliers, or anomalous events, in the temporal data.
To run these experiments, we built a module combining Snap.py \cite{snappy} and CVXPY \cite{cvxpy}. The network is stored as a Snap.py structure, and the $x$-updates of ADMM are run in parallel using CVXPY.
Even though this algorithm is capable of being distributed across many machines, we instead distribute it across multiple cores of a single machine for our prototype.
Our network-based convex optimization solver is available at \url{http://snap.stanford.edu/snapvx}, and the code for this paper can be found on the SnapVX website.

\subsection{Network-Enhanced Classification}

We first analyze a synthetic network in which each node has a support vector machine (SVM) classifier \cite{CV:95}, but does not have enough training data to accurately estimate it. The clustering of the nodes in the network occurs because some of the nodes have common underlying SVMs. The hope is that nodes can, in essence, ``borrow'' training examples from their relevant neighbors to improve their own results. Of course, neighbors with different underlying models will provide misleading information to each other. These are the edges whose lasso penalties should be non-zero, yielding different solutions at the two connected nodes. 

\xhdr{Dataset}
We randomly generate a dataset containing 1000 nodes, each with its own classifier, a support vector machine in $\reals^{50}$. Given an input $w \in \reals^{50}$, each node tries to predict $y \in \{-1,1\}$, where
\begin{equation*}
  y = \mathrm{sgn}(a_i^Tw + a_{i,0} + v),
\end{equation*}
  and $v \sim \mathcal{N}(0,1)$, the noise, is independent for each data point. An SVM involves solving a convex optimization problem from a set of training examples to obtain
$x_i =     \begin{bmatrix}
        a_i^T & a_{i,0}
    \end{bmatrix}^T \in \reals^{51}$. 
This defines a separating hyperplane to determine how to classify new inputs.
There is no way to counter the noise $v$, but an accurate $x_i$ can help us predict $y$ from $w$ reasonably accurately. Each node determines its own optimal classifier from a training set consisting of 25 $(w,y)$-pairs per node, which are used to solve for $x$. All elements in $w$, $a$, and $v$ are drawn independently from a normal distribution, 
with the $y$ values dependent on the other variables.

\xhdr{Network}
The 1000 nodes are split into 20 equally-sized groups.
Each group has a common underlying classifier, $\begin{bmatrix}
        a^T & a_0
    \end{bmatrix}^T$, while different groups have independent $a$'s. If $i$ and $j$ are in the same group, they have an edge with probability 0.5, and if they are in different groups, there is an edge with probability 0.01. Overall, this leads to a total of 17079 edges, with 28.12\% of the edges connecting nodes in different underlying groups. Even though this is a synthetic example, there are a large number of misleading edges, and each node has only 25 examples to train an SVM in $\reals^{50}$, so solving this problem is far from an easy task.

\xhdr{Optimization Parameter and Objective Function}
At node $i$, the optimization parameter $x_i = \begin{bmatrix}
        x_{i,a}^T & x_{i,0}
    \end{bmatrix}^T = \begin{bmatrix}
        a_i^T & a_{i,0}  
    \end{bmatrix}^T$ defines our estimate for the separating hyperplane for the SVM \cite{HRTZ:04}. The node then solves its own optimization problem, using its 25 training examples. At each node, $f_i$ is defined as
\begin{equation*} 
    \begin{array}{ll}
      \mbox{minimize}   & \frac{1}{2}\|x_{i,a}\|_2^2\  + \sum\limits_{i=1}^{25} c \| \varepsilon_i \|_1 \\
      \mbox{subject to} & y^{(i)}(x_{i,a}^Tw^{(i)} + x_{i,0}) \geq 1 - \varepsilon_i, \quad i = 1,\ldots,25.
  \end{array}
\end{equation*}
The $\varepsilon_i$'s are (local) slack variables. They allow points to be misclassified in the training set of a soft margin SVM \cite{CV:95}. We set $c$, the threshold parameter, to a constant which was empirically found to perform well on a common model. We solve for 51 + 25 = 76 variables at each node, so the total problem has 76,000 unknowns.

\begin{figure}[t]
\centering
  \subfigure[Convex]{\label{fig:cvxregpath}\includegraphics[width=0.23\textwidth]{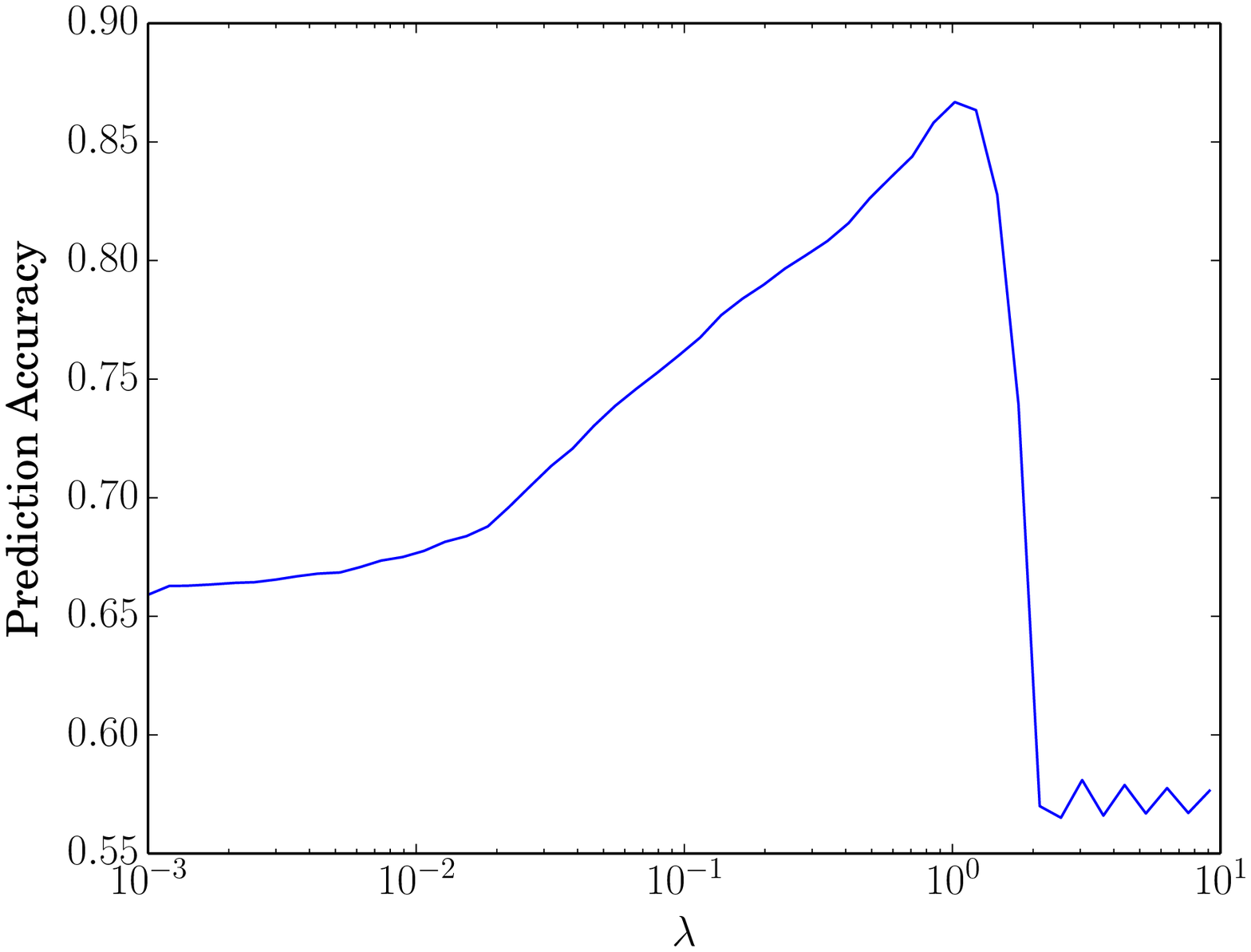}}
  \subfigure[Non-Convex]{\label{fig:noncvxregpath}\includegraphics[width=0.23\textwidth]{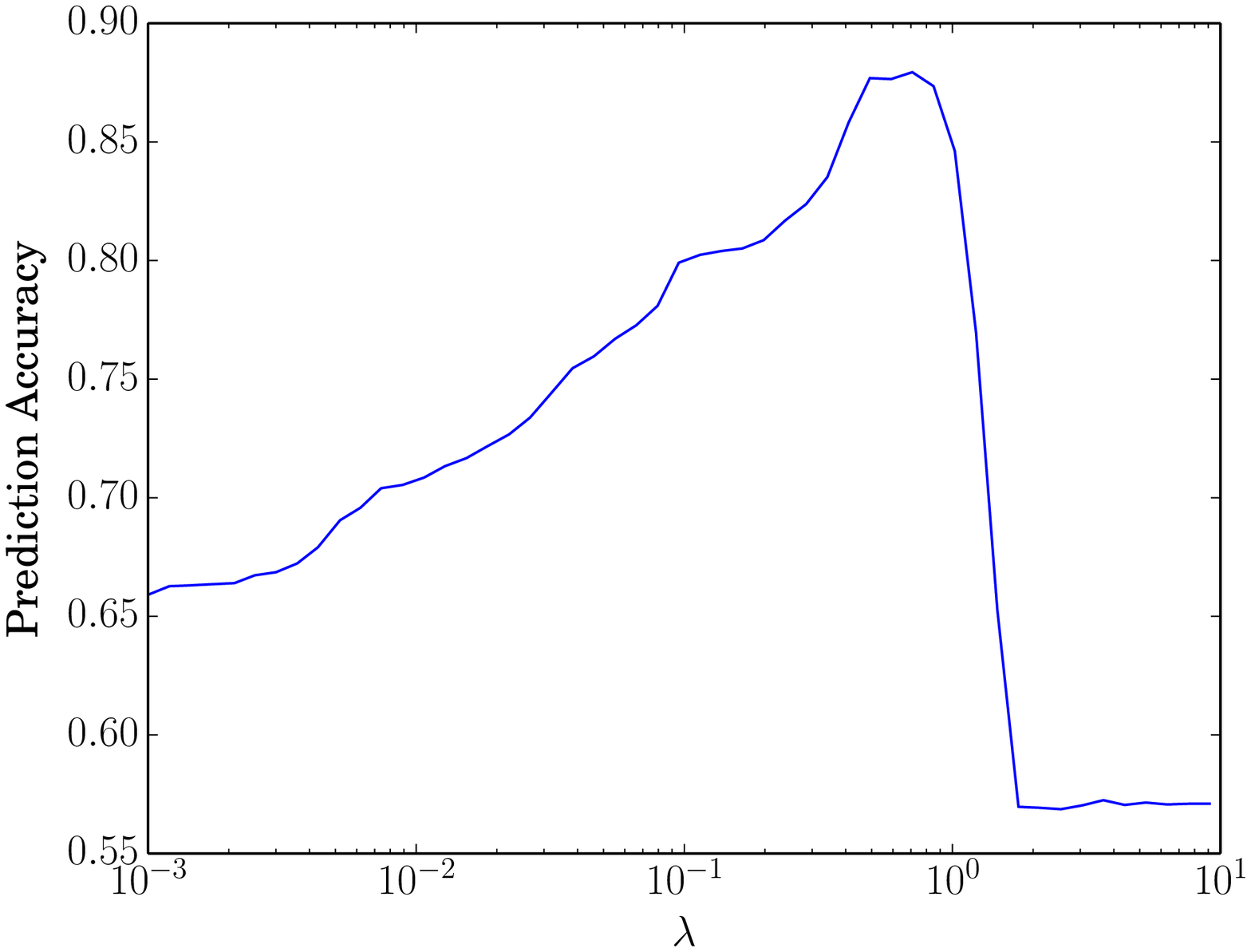}}
  \vspace{-3mm}
   \caption{SVM regularization path.}
   \vspace{-1mm}
   \label{fig:svmregpath}
\end{figure}

\xhdr{Results}
To evaluate performance, we find prediction accuracy on a separate test set of 10,000 examples (10 per node). In Figure \ref{fig:svmregpath}, we plot percentage of correct predictions vs. $\lambda$, where $\lambda$ is displayed in log-scale, over the regularization path. 
Note that the two extremes of the path represent important baselines. 

At $\lambda = 0$, each node only uses its own training examples, ignoring all the information provided by its neighbors. This is just a local SVM, with only 25 training examples to estimate a 51-dimensional vector. This leads to a prediction accuracy of 65.9\% on the test set. When $\lambda \geq \lambda_\mathrm{critical}$, the problem finds a common $x$, which is equivalent to solving a global SVM over the entire network. This assumes the entire graph is coupled together and does not allow for any edges to break. This common hyperplane at every node yields an accuracy of 57.1\%, which is barely an improvement over random guessing.
In contrast, both the convex and non-convex cases perform much better for $\lambda$'s in the middle. From Figure \ref{fig:svmregpath}, we see a distinct shape in the regularization paths. As $\lambda$ increases, the accuracy steadily improves, until a peak near $\lambda = 1$. Intuitively, this represents the point where the algorithm has approximately split the nodes into their correct clusters, each with its own classifier. As $\lambda$ continues to increase, there is a rapid drop off in performance, due to the different clusters ``pulling'' each other together. The maximum prediction accuracies on the test sets are 86.68\% (convex) and 87.94\% (non-convex). These prediction results are summarized in Table \ref{svmresults}.

\begin{table}[t]
\centering
\small
\resizebox{8cm}{!} {
    \begin{tabular}{lccrrrr}
    \hline
    Method & Maximum Prediction Accuracy \\ \hline 
    Local SVM ($\lambda = 0$) & 65.90\%\\ 
    Global SVM ($\lambda \geq \lambda_{\mathrm{critical}}$) & 57.10\%\\ 
    Convex Network Lasso & 86.68\%\\ 
    Non-Convex Network Lasso & 87.94\% \\ \hline
    \end{tabular}
    }
    \vspace{-1mm}
    \caption{SVM test set prediction accuracy.}
    \label{svmresults}
  \vspace{-1mm}
\end{table}

\begin{figure}
\centering
\includegraphics[width=0.85\linewidth]{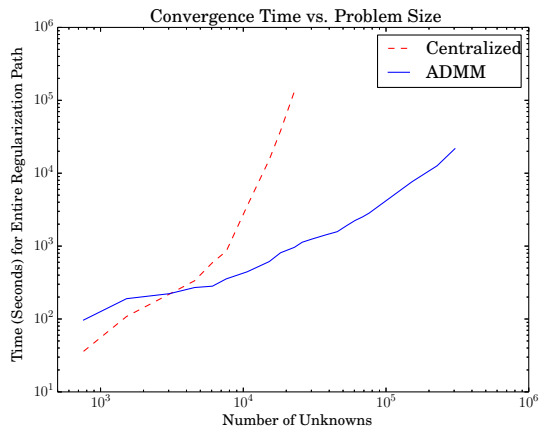}
\vspace{-3mm}
 \caption{Convergence comparison between centralized and ADMM methods for SVM problem.}
\vspace{-3mm}
 \label{fig:timingResults}
\end{figure}

\xhdr{Timing Results}
We compare our network lasso algorithm to a standard centralized method on a single 40-core CPU where the entire problem fits into memory. For the centralized case, we used the same solver (CVXPY) as in the $x$-updates for ADMM. While wrapped in a Python layer, CVXPY's underlying solver uses ECOS \cite{ecos},
an open-source software package specifically designed for high performance numerical optimization, so the Python overhead is negligible when it comes to the cost of scaling to large problems. We show the results on the synthetic SVM example to scale the problem size over several orders of magnitude. 
We solve the problem at 12 geometrically spaced values of $\lambda$ to span the entire regularization path.  
We use $\frac{n}{20}$ underlying SVM clusters, where $n$ is the number of nodes.
The entire regularization path is one large problem (consisting of 12 smaller ones), and we measure its total runtime. Note that each node in this case is solving its own SVM, with additional coupling constraints due to the network lasso on the edges. We vary the total number of nodes, and the results are shown in Figure \ref{fig:timingResults}. We see that, in this example, the centralized method scales on the order of problem size cubed, whereas ADMM takes closer to linear time, until other concerns such as memory limitations begin to factor in. By the time there are 20,000 unknowns, ADMM is already 100 times faster, and this discrepancy in convergence time only grows as the problem gets larger. 


To further test our algorithm, we also solve a larger yet simpler problem. We build a random 3-regular graph (every node has a degree of 3) with 2000 nodes. The objective function at each node is $f_i(x_i) = \| x_i - a_i\|_2^2$, where $a_i$ is a random vector in $\reals^{q}$. We can modify the value of $q$ to vary the total number of unknowns. 
We pick a single (constant) $\lambda$ in the middle of the regularization path and see how long it takes to solve the problem using ADMM. The results are shown in Table \ref{largeTiming}. We can compute a solution for 1 million unknowns in seconds, and for 100 million in under 15 minutes.
It is worth reiterating that at each step, at each node, we use CVXPY rather than a more specialized solver for the x-update subproblem. This allows the same solver to work on any convex node objective, rather than being constrained to specific classes of functions, and yet it is still able to scale to tens of millions of unknown variables.



\begin{table}[t]
\centering
\small
\resizebox{8cm}{!} {
    \begin{tabular}{l r}
    \hline
    Number of Unknowns & ADMM Solution Time (seconds) \\ \hline
    100,000 & 12.20 \\ 
    1 million & 18.16\\  
    10 million & 128.98\\ 
    100 million & 822.62\\ \hline 
    \end{tabular}
    }
    \vspace{-1mm}
    \caption{Convergence time for large-scale 3-regular graph solved at a single (constant) value of $\lambda$.}
    \label{largeTiming}
  \vspace{-1mm}
\end{table}


\subsection{Spatial Clustering with Regressors}

In this example, as described in the introduction, we attempt to estimate the price of homes based on latitude/longitude data and a set of features. Home prices often cluster together along neighborhood lines.
In this case, the clustering occurs when nearby houses have similar pricing models, while edges that have non-zero edge differences will be between those in different neighborhoods. As houses are grouped together, each cluster builds its own local linear regression model to predict prices in its region. Then, when there is a new house, we can infer its regression model from the local neighborhood to estimate the sales price.

\xhdr{Dataset} We look at a list of real estate transactions over a one-week period in May 2008 in the Greater Sacramento area\footnote{Data available at \url{http://support.spatialkey.com/spatialkey-sample-csv-data/}.}. This dataset contains information on 985 sales, including latitude, longitude, number of bedrooms, number of bathrooms, square feet, and sales price. However, as often happens with real data, we are missing some of the values. 
17\% of the home sales are missing at least one of the features; \ie, some of the bedroom/bathroom/size data is not provided.
The price and all attributes are standardized to zero mean and unit variance, so any missing features are ignored by setting the value to zero, the average.
To verify our results, we use a random subset of 200 houses as our test set.

\xhdr{Network} We build the graph by using the latitude/longitude coordinates of each house. After removing the test set, we connect every remaining house to the five nearest homes with an edge weight inversely proportional to the distance between the houses. 
If house $j$ is in the set of nearest neighbors of $i$, there is an undirected edge regardless of whether or not house $i$ is one of $j$'s nearest neighbors. The resulting graph leaves 785 nodes, 2447 edges, and has a diameter of 61.

\xhdr{Optimization Parameter and Objective Function}
At each node, we solve for
\begin{equation*} 
    x_i =     \begin{bmatrix}
        a_i & b_i & c_i & d_i
     \end{bmatrix}^T,
\end{equation*}
which gives us the weights of the regressors. The price estimate is given by 
\begin{equation*} 
    \overline{\mathrm{price}_i} = a_i \cdot \mathrm{Bedrooms} + b_i \cdot \mathrm{Bathrooms} + c_i \cdot \mathrm{SQFT} + d_i,
\end{equation*}
where the constant offset $d_i$ is the ``baseline''. To prevent overfitting, we regularize the $a_i$, $b_i$, and $c_i$ terms, everything besides the offset. The objective function at each node then becomes
\begin{equation*} 
    \begin{array}{ll}
       f_i = \|\overline{\mathrm{price}_i} - \mathrm{price}_i\|_2^2 + \mu \left\lVert \tilde{x_i} \right\rVert_2^2
    \end{array}
\end{equation*}
where $\tilde{x_i} =    \begin{bmatrix}
        a_i & b_i & c_i
     \end{bmatrix}^T$, $\mathrm{price}_i$ is the actual sales price, and $\mu$ is a constant regularization parameter.

To predict the prices on the test set, we connect each new house to the 5 nearest homes, weighted by inverse distance, just like before. We then infer the value of $x_j$ at node $j$ by solving problem \eqref{inference}, and we use this value to estimate the sales price.

\begin{figure}[!t]
\centering
  \subfigure[Convex]{\label{fig:cvxregpathHousing}\includegraphics[width=0.49\linewidth]{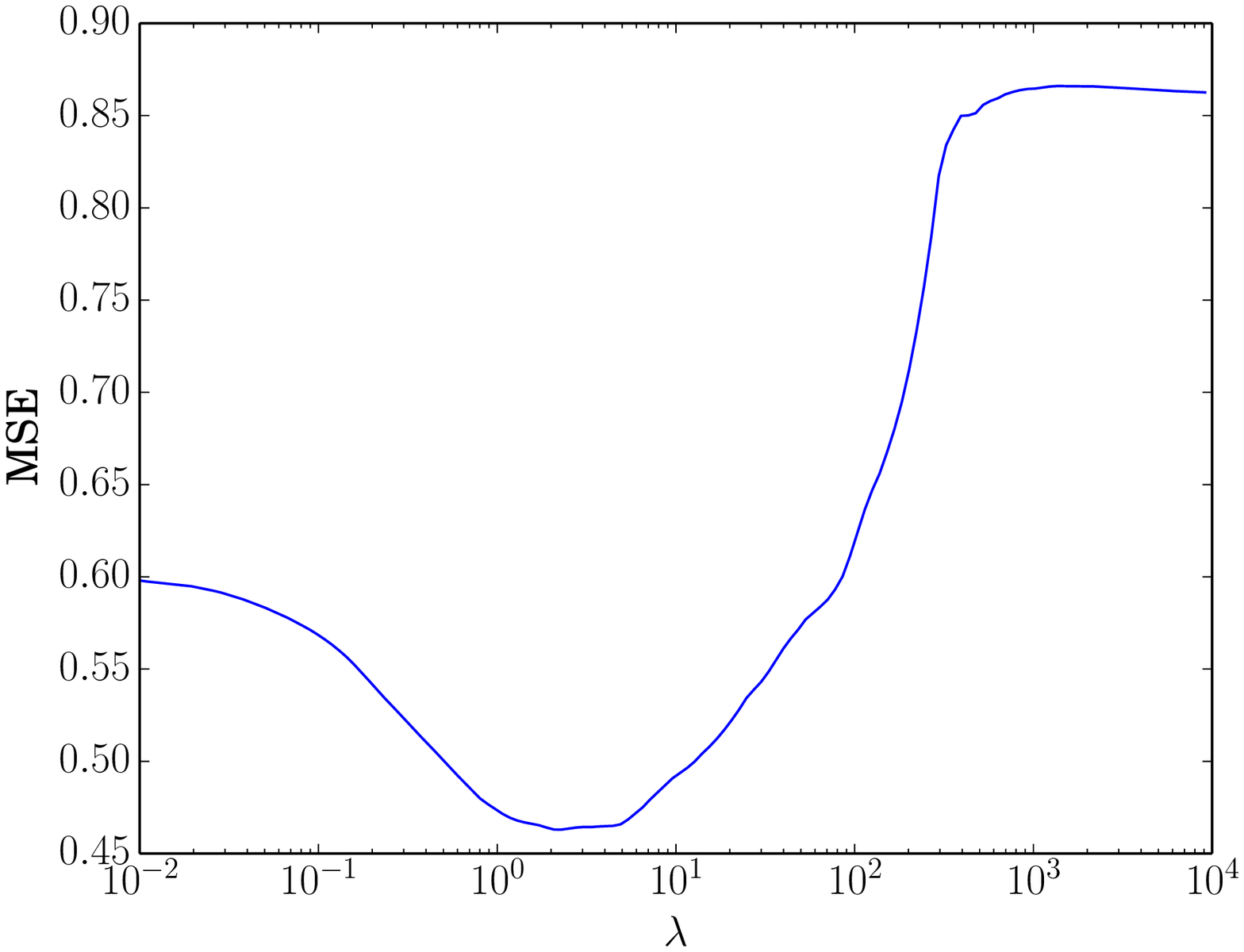}}
  \subfigure[Non-Convex]{\label{fig:noncvxregpathHousing}\includegraphics[width=0.49\linewidth]{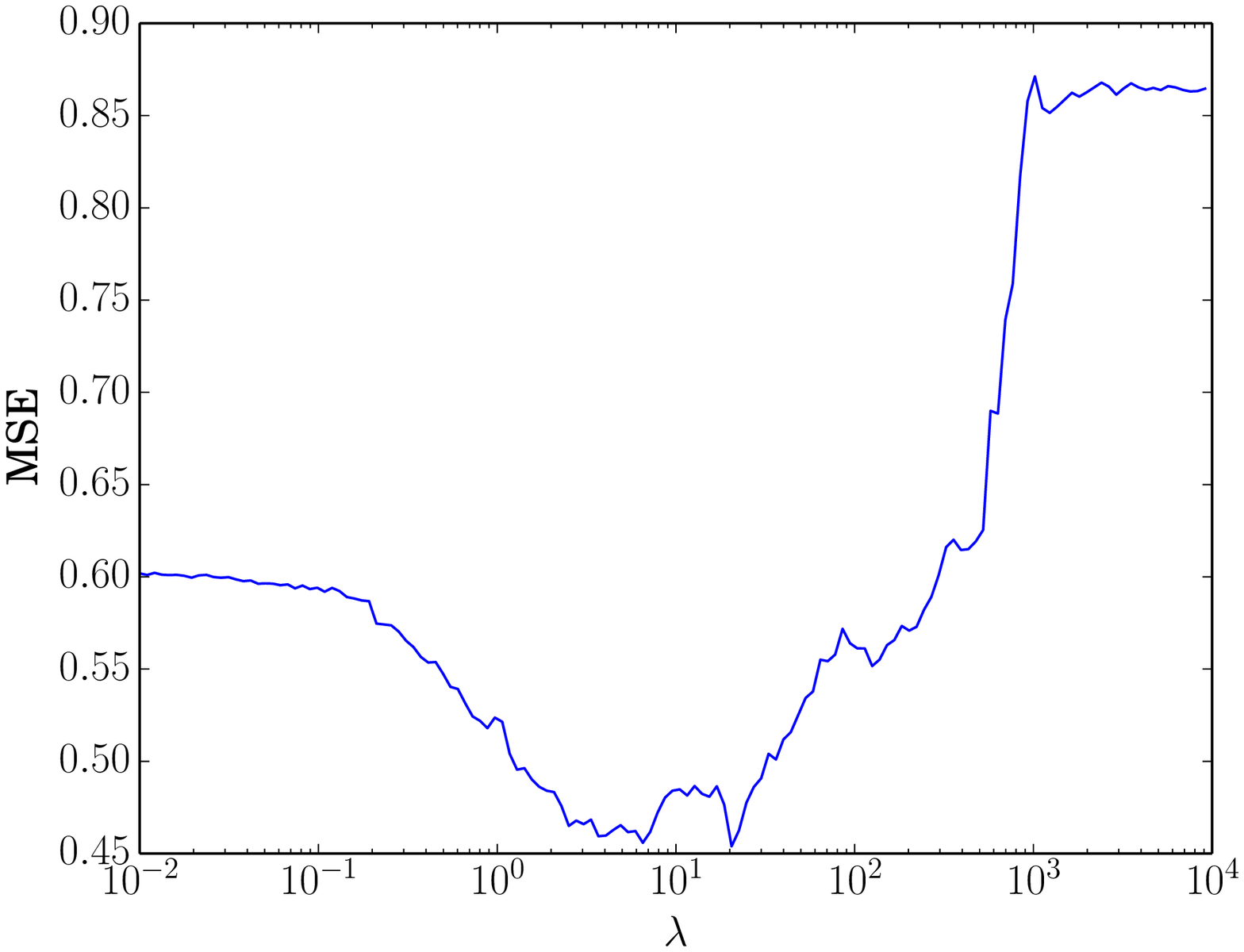}}
  \vspace{-3mm}
   \caption{Regularization path for housing data.}
   \vspace{-1mm}
   \label{fig:regpathHousing}
\end{figure}

\xhdr{Results} 
We plot the mean squared error (MSE) vs. $\lambda$ in Figure \ref{fig:regpathHousing} for both the convex and non-convex formulations of the problem. Once again, the two extremes of the regularization path are relevant baselines.
\newcolumntype{d}[1]{D{.}{.}{4} }
\begin{table}[!t]
\centering
\small
\resizebox{8cm}{!} {
    \begin{tabular}{l d{1}}
    \hline
    Method & \multicolumn{1}{r}{Mean Squared Error (MSE)} \\ \hline 
    Geographic ($\lambda = 0$) & 0.6013\\ 
    Regularized Linear Regression ($\lambda \geq \lambda_{\mathrm{critical}}$) & 0.8611\\ 
    Naive Prediction (Global Mean) & 1.0245\\ 
    Convex Network Lasso & 0.4630\\ 
    Non-Convex Network Lasso & 0.4539 \\ \hline 
    \end{tabular}
    }
    \vspace{-1mm}
    \caption{MSE for housing price predictions on test set.}
    \label{housingresults}
  \vspace{-1mm}
\end{table}

At $\lambda = 0$, the regularization term in $f_i(x_i)$ insures that the only non-zero element of $x_i$ is $d_i$. This ignores the regressors and is a prediction based solely on spatial data. Our estimate for each new house is simply the weighted median price of the 5 nearest homes, which leads to an MSE of 0.6013 on the test set.  For large $\lambda$'s, we are fitting a common model for all the houses. This is just regularized linear regression on the entire dataset and is the canonical method of estimating housing prices from a series of features. Note that this approach completely ignores the geographic network. As expected, it performs rather poorly, with an MSE of 0.8611. Since the prices are standardized with unit variance, a naive guess (with no information about the house) would just be the global average of the training set, which has an MSE of 1.0245. The convex and non-convex methods are both maximized around $\lambda = 5$, with minimum MSE's of 0.4630 and 0.4539, respectively. 

\begin{figure}[!t]
\centering 
  \subfigure[$\lambda$ = 0.1]{\label{fig:heatmap1}\includegraphics[width=0.49\linewidth]{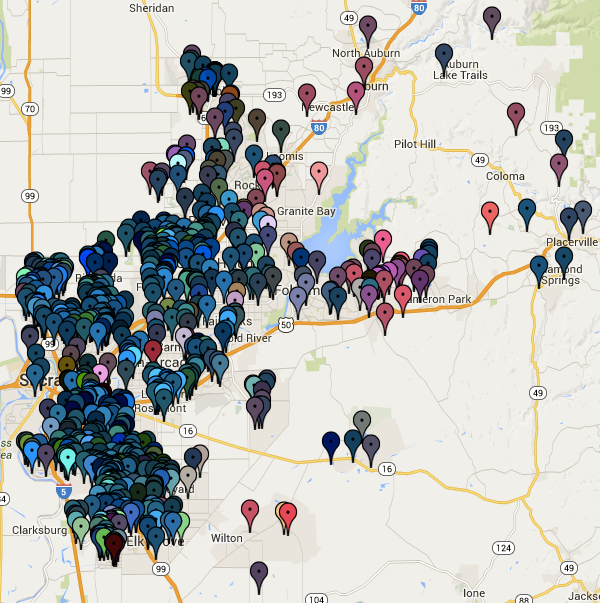}}
  \subfigure[$\lambda$ = 1000]{\label{fig:heatmap2}\includegraphics[width=0.49\linewidth]{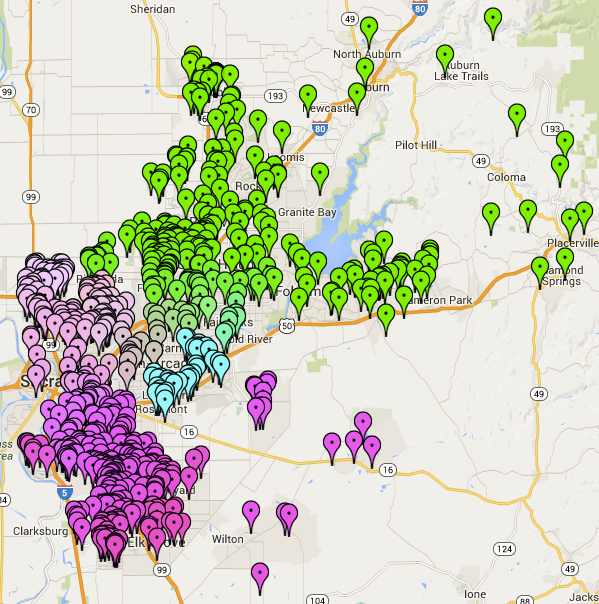}}
  \subfigure[$\lambda$ = 10] {\label{fig:heatmap3}\includegraphics[width=0.99\linewidth]{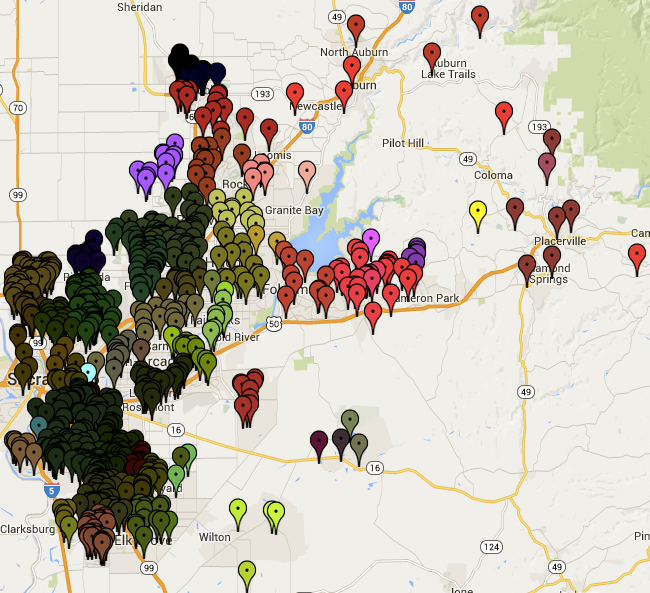}}
  \vspace{-3mm}
   \caption{Regularization path clustering pattern.}
   \vspace{-1mm}
   \label{fig:heatmap}
\end{figure}

We can visualize the clustering pattern by overlaying the network on a map of Sacramento. We plot each sale with a marker, colored according to its corresponding $x_i$ (so houses with similar colors have similar models, and those with the same color are in consensus). With this, we see how the clustering pattern emerges. In Figure \ref{fig:heatmap}, we look at this plot for three values of $\lambda$. In \ref{fig:heatmap1}, $\lambda$ is too small, so the neighborhoods have not yet formed. On the other hand, in \ref{fig:heatmap2}, $\lambda$ is too large. The clustering is clear, but it performs poorly because it forces together neighborhoods which are very different. Figure \ref{fig:heatmap3} is a viable choice of $\lambda$, leading to low MSE while showing a clear partitioning of the network into neighborhoods of different sizes.

Aside from outperforming the baselines, this method is also well-suited to detect and handle anomalies. As shown in the plots, outliers are often treated as single-element clusters, for example the yellow house on the right side of \ref{fig:heatmap3}. These houses are ones which do not fit in with their local model (for a variety of possible reasons), but using the network lasso, neither they nor their neighbors are adversely affected too significantly by each other. Of course, as $\lambda$ approaches $\lambda_{\mathrm{critical}}$, these clusters are forced together into consensus. However, near the optimal $\lambda$, we accurately classify these anomalies, isolate them from the rest of the graph, and build separate and relatively accurate models for both subsets.

\subsection{Event Detection in Time Series Data}

Lastly, we aim to predict the existence of certain ``events'' in a building, those which were officially listed by the building coordinator. We are given the entry and exit data from the building over a 15 week interval. For these events, we expect to see an anomalous increase in traffic. Note that this is just a partial ground truth, only containing events officially reported by the coordinator, and many unreported events likely occurred during this interval. Therefore, ``false positives'' are not necessarily incorrect, so the absolute results (how accurately we predict the events) are not a perfect indicator of performance. However, this provides a good benchmark, especially when compared to a common baseline.

\xhdr{Dataset}
The data comes from the main door of the Calit2 building at UC Irvine. This count data, the number of entries and exits, is reported once every 30 minutes over the course of 15 weeks from July to November 2005, for a total of 5,040 readings\footnote{Data from \url{https://archive.ics.uci.edu/ml/datasets/CalIt2+Building+People+Counts} \cite{BL:13}.}. Additionally, we use a list of the 30 official events which occurred inside the building during that interval.

\xhdr{Network}
We build a linear network where node $i$, covering the $i$th interval in the time series, has only two edges. These connect it to nodes $i-1$ and $i+1$. The first and last nodes only have one edge, leaving 5,040 nodes and 5,039 edges. There are more complicated ways to model the coupling of time series data, but we opt for simplicity since our goal is to show one approach, rather than necessarily the optimal method, of solving this class of problems.




\xhdr{Optimization Parameter and Objective Function}
Traffic is periodic on a weekly basis. That is, a relatively similar number of people enter and exit the building on, for example, Mondays from 1:00 - 1:30PM. We do not care for instance that there is more traffic at 1:00 PM than at 1:00 AM. This is not an indicator that an event occurred at 1PM. Instead, we care about the number of people relative to the periodic signal. We let 
\begin{equation*} 
    \overline{x}_i =     \begin{bmatrix}
        \mathrm{in}_i - \overline{\mathrm{in}}(i \bmod 336) \\[0.3em]
        \mathrm{out}_i - \overline{\mathrm{out}}(i \bmod 336)
     \end{bmatrix},
\end{equation*}
where $\overline{\mathrm{in}}(i \bmod 336)$ and $\overline{\mathrm{out}}(i \bmod 336)$ are the median value of entrances/exits for the given time and day of the week ($7 \cdot 24 \cdot 2 = 336$) over the 15 week interval. We use the median because the mean can be skewed by the increases due to actual events.

The objective function is defined as
\begin{equation*} 
    f_i = \|x_i - \overline{x}_i\|_2^2 + \mu \|x_i\|_2.
\end{equation*}
The variable that we optimize over, $x_i$, is an attempt to match the non-periodic signal at that time. The regularization term on $x_i$ is a lasso penalty, so only a select few of the $x$'s will be non-zero. These non-zero values refer to the times of the anomalous events that we are trying to predict. It is worth noting that for any finite network lasso parameter $\lambda$, there exists a $\mu$ large enough so that every $x_i$ is guaranteed to be $[0, 0]^T$.

An event often manifests itself as a sustained period of increased activity. Therefore, we declare an event on the interval $[t, t+k]$ if 
\begin{equation*} 
    x_{i,\mathrm{in}} + x_{i,\mathrm{out}} > 0 \quad i \in [t, t+k].
\end{equation*}
We vary $\mu$ to change the number of events predicted. For small $\mu$, the slightest noise can be interpreted as an event. Large $\mu$'s lead to fewer predictions, until eventually every $x(t)$ is forced to 0, as mentioned before. The parameter $\lambda$ determines the average event length, as it encourages prolonged increases in activity and discourages single outliers from being picked up. However, in this example, the model is relatively robust to changes in $\lambda$ (up to a certain point), so we keep it constant as we vary $\mu$, as a slight modification of the regularization path from previous experiments.

\xhdr{Baseline}
This type of problem is often modeled as a Poisson process, so we use that as our baseline method \cite{IHS:06}. We consider each time and day of the week as having an independent Poisson rate $\lambda$ (which is unrelated to the regularization parameter with the same name in the network lasso). We set $\lambda$, the ``expected'' number of count data, to the maximum likelihood estimate of a Poisson process, the mean of the 15 values. $\lambda_{\mathrm{in}}$ and $\lambda_{\mathrm{out}}$ are calculated independently. We define an event from $[t, t+k]$ if
\begin{align*} 
    P(N(i), \lambda(i)) &= \left(\frac{e^{-\lambda_{\mathrm{in}}}\lambda_{\mathrm{in}}^{N_{\mathrm{in}}(i)}}{N_{\mathrm{in}}(i)!}\right) \left(\frac{e^{-\lambda_{\mathrm{out}}}\lambda_{\mathrm{out}}^{N_{\mathrm{out}}(i)}}{N_{\mathrm{out}}(i)!}\right) \\
    &< \epsilon \quad \quad i \in [t, t+k].
\end{align*}
This says that the given number of entries and exits at time $i$ occurs with probability less than $\epsilon$. Since only large totals should trigger a predicted event (rather than abnormally low entry/exit numbers), one final requirement is that either $N_{in} > \lambda_{in}$ or $N_{out} > \lambda_{out}$ for every $t$ in the interval. Varying the threshold $\epsilon$, similar to $\mu$ for our approach, changes the number of predicted events.

\xhdr{Results}
For both our model and the baseline, we compute the number of correct events vs. number of predicted events. We define a correct prediction as one in which the prediction and the true event overlap. The accuracy of all three approaches at several key points is summarized in Table \ref{anomalyresults}. As shown, both the convex and non-convex methods outperform the Poisson baseline (though the convex approach does noticeably better than the non-convex). The Poisson is able to catch the ``low-hanging fruit'', the easy-to-detect events, with relatively good accuracy. The discrepancy arises in the less obvious ones. Again, this is just a partial ground truth and it is likely that there are many more than 30 events, but the poor performance of the Poisson method --- it takes 264 predictions to find all 30 events --- suggests that it may be an imperfect method of event detection.
\begin{table}[!t]
\centering
\small
\resizebox{8cm}{!} {
    \begin{tabular}{l|rrrr}
    \hline
    Number of Correct Events Detected & \multicolumn{3}{c}{Predicted Events}  \\
    & Convex & Non-Convex & Poisson \\ \hline 
    30 & 146 & 201 & 264 \\ 
    29 & 125 & 135 & 214 \\ 
    28 & 116 & 121 & 201 \\ 
    27 & 101 & 116 & 188 \\ 
    26 & 97 & 114 & 131 \\ 
    24 & 76 & 78 & 100 \\ 
    18 & 56 & 64 & 62 \\ \hline
    \end{tabular}
    }
    \vspace{-1mm}
    \caption{Number of required predictions to detect events.}
    \label{anomalyresults}
  \vspace{-1mm}
\end{table}
Note that more complicated models, specifically tuned for outlier detection, may beat these results. For example when an event occurs, we expect to see a large spike in inbound traffic at the beginning of the event, and a similar outbound one at the end. Our approach could easily be modified in future work to account for additional information such as this. However, as a simple model and a proof of concept, these results are very encouraging.

%% file: 080conclusion.tex
In this paper, we have shown that within one single framework, it is possible to better understand and improve on many common machine learning and network analysis problems. The network lasso is a useful way of representing convex optimization problems, and the magnitude of the improvements in the experiments show that this approach is worth exploring further, as there are many potential ideas to build on. The non-convex method gave comparable performance to the convex approach, and we leave for future work the analysis of different non-convex functions $\phi(u)$. It is also possible to look at the sensitivity of these results to the structure of the network. For example, we could attempt to iteratively reweigh the edge weights to attain some desired outcome. 
Within the ADMM algorithm, there are many ways to improve speed, performance, and robustness. This includes finding closed-form solutions for common objective functions $f_i(x_i)$, automatically determining the optimal ADMM parameter $\rho$, and even allowing edge objective functions $f_e(x_i, x_j)$ beyond just the weighted network lasso. As this topic develops further, there is an opportunity for easy-to-use software packages which allow programmers to solve these types of large-scale optimization problems in a distributed setting without having to specify the implementation details, which would greatly improve the practical benefit of this work. 

\subsection*{Acknowledgments} 
The authors would like to thank Trevor Hastie for his advice on the network lasso, Stephen Bach and Christopher R\'e for their help with graphical models, and Rok Sosi\v{c} for his assistance during the large-scale implementation. This research has been supported in part by the Sequoia Capital Stanford Graduate Fellowship,
NSF
IIS-1016909,              
CNS-1010921,              
IIS-1149837,              
IIS-1159679,              
ARO MURI,                 
DARPA XDATA, SMISC, SIMPLEX,
Stanford Data Science Initiative,
Boeing,                    
Facebook,
Volkswagen,
and Yahoo.

%% file: 100appendixB.tex
We will show that the solution to
\begin{equation*}
    \begin{array}{ll}
     \mbox{minimize}  &\lambda w_{ij}\|z_{ij} - z_{ji}\|_2 + (\rho/2) \bigl(  \|x_i^{k+1} - z_{ij} + u_{ij}^k\|_2^2 + \\
     &\|x_j^{k+1} - z_{ji} + u_{ji}^k\|_2^2 \bigr),
    \end{array}
    \end{equation*}
with variables $z_{ij}$ and $z_{ji}$, is
        \begin{align*} 
     z_{ij}^\star &= \theta(x_i + u_{ij}) + (1-\theta)(x_j + u_{ji}) \\
        z_{ji}^\star &= (1-\theta)(x_i + u_{ij}) + \theta(x_j + u_{ji}),
        \end{align*}	
where $\theta$ is defined in equation \eqref{thetaDef}.

We first note that the objective is strictly convex, so the solution is unique. As in \S 4, we let
         \[
     	a = x_i^{k+1} + u_{ij}^k, \quad
	b = x_j^{k+1} + u_{ji}^k, \quad
	c = \lambda w_{ij},
        \]
so the original problem turns into
\begin{equation*}
    \begin{array}{ll}
     \mbox{minimize}  &c\|z_{ij} - z_{ji}\|_2 + (\rho/2) \left(  \|a - z_{ij}\|_2^2 + \|b - z_{ji}\|_2^2 \right).
    \end{array}
    \end{equation*}
There are two possible cases for the optimal values $z_{ij}^\star$ and $z_{ji}^\star$. 

\xhdr{Case 1:  $z_{ij}^\star = z_{ji}^\star$}
If the two variables are equal, then $\|z_{ij} - z_{ji}\|_2 = 0$, so the only terms remaining are 
         \begin{equation*}
	(\rho/2) \left(  \|a - z_{ij}\|_2^2 + \|b - z_{ji}\|_2^2 \right).
        \end{equation*}
Minimizing over the constraint that $z_{ij} = z_{ji}$ yields $z_{ij}^\star = z_{ji}^\star = (1/2)(a + b)$, with objective value $\rho/4 \| a - b\|^2_2$.

\xhdr{Case 2: $z_{ij}^\star \neq z_{ji}^\star$}
When the two variables are not equal, the objective is differentiable. In this case, the necessary and sufficient condition for optimality is $\nabla f = 0$, or
         \begin{equation*}
	\nabla \left(c \|z_{ij} - z_{ji}\|_2 + (\rho/2) \|a - z_{ij}\|_2^2 + (\rho/2) \|b - z_{ji}\|_2^2 \right) = 0.
        \end{equation*}
The gradient can be written as
         \begin{equation*} 
	c 
	\begin{bmatrix}
       \frac{z_{ij} - z_{ji}}{\|z_{ij} - z_{ji}\|_2 }       \\[0.3em]
       -\frac{z_{ij} - z_{ji}}{\|z_{ij} - z_{ji}\|_2 }  
     \end{bmatrix} 
     + 
     \begin{bmatrix}
       -\rho (a - z_{ij})       \\[0.3em]
       -\rho (b - z_{ji})
     \end{bmatrix}
     =      \begin{bmatrix}
       0       \\[0.3em]
       0
     \end{bmatrix},
        \end{equation*}
so the two equations that must be satisfied are
         \begin{equation*}
     	c \frac{z_{ij} - z_{ji}}{\|z_{ij} - z_{ji}\|_2 }  - \rho (a - z_{ij}) = 0, \;\;\;  \;\;\; -c \frac{z_{ij} - z_{ji}}{\|z_{ij} - z_{ji}\|_2 }  - \rho (b - z_{ji}) = 0.
        \end{equation*}
Letting $\mu = \|z_{ij} - z_{ji}\|_2$, we get
         \begin{equation*}
     	c (z_{ij} - z_{ji}) = \mu \rho (a - z_{ij}), \;\;\;\;  \;\;\;\; -c (z_{ij} - z_{ji}) = \mu \rho (b - z_{ji}).
        \end{equation*}
Adding the two equations gives
         \begin{equation*}
     	z_{ij} + z_{ji} = a + b,
        \end{equation*}
and subtracting them leads to
        \begin{equation*}
	z_{ij} - z_{ji} = \frac{\mu \rho (a - b)}{2c + \mu \rho}.
        \end{equation*}   
Treating $\mu$ as a constant, this yields a system of linear equations for $z_{ij}$ and $z_{ji}$, which we solve to obtain
        \begin{equation*}
	 z_{ij} = \theta a + (1-\theta)b, \quad \quad z_{ji}= (1-\theta)a + \theta b,
        \end{equation*}
where          
	\begin{equation*} 
     \theta =  \frac{1}{2} +  \frac{\mu \rho}{4c + 2\mu \rho}.
        \end{equation*}
We know that $\mu = \|z_{ij} - z_{ji}\|_2$, so we plug in for $z_{ij}$ and $z_{ji}$,
         \begin{equation*}
	\mu = \|z_{ij} - z_{ji}\|_2 = \left\|\frac{\mu \rho (a - b)}{2c + \mu \rho}\right\|_2 = \frac{\mu \rho}{2c + \mu \rho} \| a - b \|_2,
	\end{equation*}      
which reduces to
  	\begin{equation*}
	1 = \frac{\rho}{2c + \mu \rho} \| a - b \|_2.
	\end{equation*}      
 From this, we can solve for $\mu$,
        \begin{equation*}
	\mu = \| a - b \|_2 - \frac{2c}{\rho}.
        \end{equation*}
We plug in $\mu$ to solve for $\theta$, which yields
         \begin{equation*}
	\theta = \frac{1}{2} +  \frac{\left(\| a - b \|_2 - \frac{2c}{\rho}\right) \rho}{4c + 2 \rho\left(\| a - b \|_2 - \frac{2c}{\rho}\right)} = \frac{1}{2} +  \frac{\rho \| a - b \|_2 - 2c}{2\rho \| a - b \|_2}.
	\end{equation*}
 This is then reduced to
        \begin{equation*}
	\theta = 1 - \frac{c}{\rho \|a - b\|_2}.
        \end{equation*}
However, this only holds if $z_{ij} \neq z_{ji}$. When this condition is not satisfied, we know the solution is case 1, which is equivalent to $\theta = \frac{1}{2}$. When it is satisfied, we need to compare the resulting objective with $\rho/4 \| a - b\|^2_2$, the value from case 1. Routine calculations show that this holds when $\theta > \frac{1}{2}$. Therefore, combining these equations and plugging in for $a$, $b$, and $c$, we arrive at our solution,
         \begin{equation*}
     \theta = \mathrm{max}\left(1 - \frac{\lambda w_{ij}}{\rho \|x_i + u_{ij} - (x_j + u_{ji})\|_2}, 0.5 \right). 
        \end{equation*}